\documentclass[sigconf,10pt]{acmart}

\usepackage{times}
\usepackage{tikz}
\usepackage{amsmath}

\usepackage{booktabs} 
\usepackage{import}
\usepackage{graphicx}
\usepackage{epstopdf}
\usepackage{amsmath}
\usepackage{amsfonts}
\usepackage{commath}
\usepackage{subfigure}
\usepackage{diagbox}
\graphicspath{{Fig/}}
\usepackage{booktabs} 
\usepackage{algorithm}
\usepackage{algorithmicx} 
\usepackage{algpseudocode}
\usepackage{paralist}
\usepackage{subcaption}
\usepackage{multirow}
\usepackage{xspace}
\usepackage{graphicx}
\usepackage{subfigure}
\usepackage{color}
\usepackage{xcolor}
\usepackage{color,soul}
\usepackage{array}
\usepackage{diagbox}
\usepackage{verbatim} 
\usepackage{colortbl}
\usepackage{upgreek}
\usepackage{enumitem}
\usepackage{gensymb} 
\usepackage{amsmath,bm}
\usepackage{graphicx}
\usepackage{subcaption}
\usepackage{adjustbox}
\usepackage{geometry}
\usepackage{url}

\definecolor{mygray}{gray}{.95}
\definecolor{mypink}{rgb}{.99,.91,.95}
\definecolor{myblue}{rgb}{.93,.94,.97}
\definecolor{mygreen}{rgb}{0,.0.65,.30}

\definecolor{ao}{rgb}{0.54, 0.17, 0.89}

\newcommand{\name}{InverTwin\xspace}

\def\ie{\textit{i.e.}\xspace}
\def\eg{\textit{e.g.}\xspace}
\def\etc{\textit{etc.}\xspace}

\def\wrt{\textit{w.r.t.}\xspace}

\newcommand{\xya}[1]{{\color{black}{#1}}}

\renewcommand\footnotetextcopyrightpermission[1]{}
\begin{document}
\settopmatter{authorsperrow=4}

\author{Xingyu Chen}
\affiliation{%
  \institution{University of California San Diego}
  \city{San Diego}
  \state{CA}
  \country{USA}
}

\author{Jianrong Ding}
\affiliation{%
  \institution{The University of Hong Kong}
  \city{Hong Kong}
  \country{China}
}

\author{Kai Zheng}
\affiliation{%
  \institution{University of California San Diego}
  \city{San Diego}
  \state{CA}
  \country{USA}
}

\author{Xinmin Fang}
\affiliation{%
  \institution{University of Colorado Denver}
  \city{Denver}
  \state{CO}
  \country{USA}
}

\author{Xinyu Zhang}
\affiliation{%
  \institution{University of California San Diego}
  \city{San Diego}
  \state{CA}
  \country{USA}
}

\author{Chris Xiaoxuan Lu}
\affiliation{%
  \institution{University College London}
  \city{London}
  \country{United Kingdom}
}

\author{Zhengxiong Li}
\affiliation{%
  \institution{University of Colorado Denver}
  \city{Denver}
  \state{CO}
  \country{USA}
}
\renewcommand{\shortauthors}{X. Chen et al.}

\title[\name]{\name:  Solving Inverse Problems via \\ Differentiable Radio Frequency Digital Twin 
}

\begin{abstract}

Digital twins (DTs), virtual simulated replicas of physical scenes, are transforming various industries. However, their potential in radio frequency (RF) sensing applications has been limited by the unidirectional nature of conventional RF simulators. In this paper, we present \name, an optimization-driven framework that creates RF digital twins by enabling bidirectional interaction between virtual and physical realms. \name overcomes the fundamental differentiability challenges of RF optimization problems through novel design components, including path-space differentiation to
address discontinuity in complex simulation functions, and a radar surrogate
model to mitigate local non-convexity caused by RF signal periodicity. 
These techniques enable smooth gradient propagation and robust optimization of the DT model. Our implementation and experiments demonstrate \name's versatility
and effectiveness in augmenting both data-driven and model-driven RF sensing systems for DT reconstruction.

\end{abstract}

\maketitle

\section{Introduction}

Radio Frequency (RF) sensing has emerged as a transformative technology with diverse applications across multiple domains, including healthcare monitoring, human activity recognition, indoor localization, and industrial automation. The non-invasive nature, ability to penetrate non-metallic materials, and sensitivity to subtle environmental changes position RF sensing as a cornerstone in the development of smart environments and Internet of Things (IoT) ecosystems. Recent advancements in RF sensing, particularly radar sensing, have demonstrated its potential in detecting micro-gestures \cite{zheng2019zero,lien2016soli,li2022towards}, monitoring vital signs \cite{xu:mobicom2017}, and even enabling through-wall imaging \cite{zhao2018rf, zhao2019through, xue2021mmmesh}, showcasing its versatility and significance in both research and practical applications.

Mainstream RF sensing approaches adopt either data-driven or model-driven
methodologies. Data-driven techniques, while powerful in extracting complex
patterns from large datasets, often struggle with generalization to unseen
scenarios and environments. This limitation stems from their inherent
dependency on the diversity of training data, which may not always capture the
full spectrum of real-world variability. Conversely, model-driven approaches,
grounded in RF propagation models and signal processing principles, offer robust theoretical foundations but frequently fall short in incorporating the implicit, experiential knowledge of the physical world that data-driven methods excel at capturing.

\begin{figure}[t]
\centering
\includegraphics[width=0.48\textwidth]{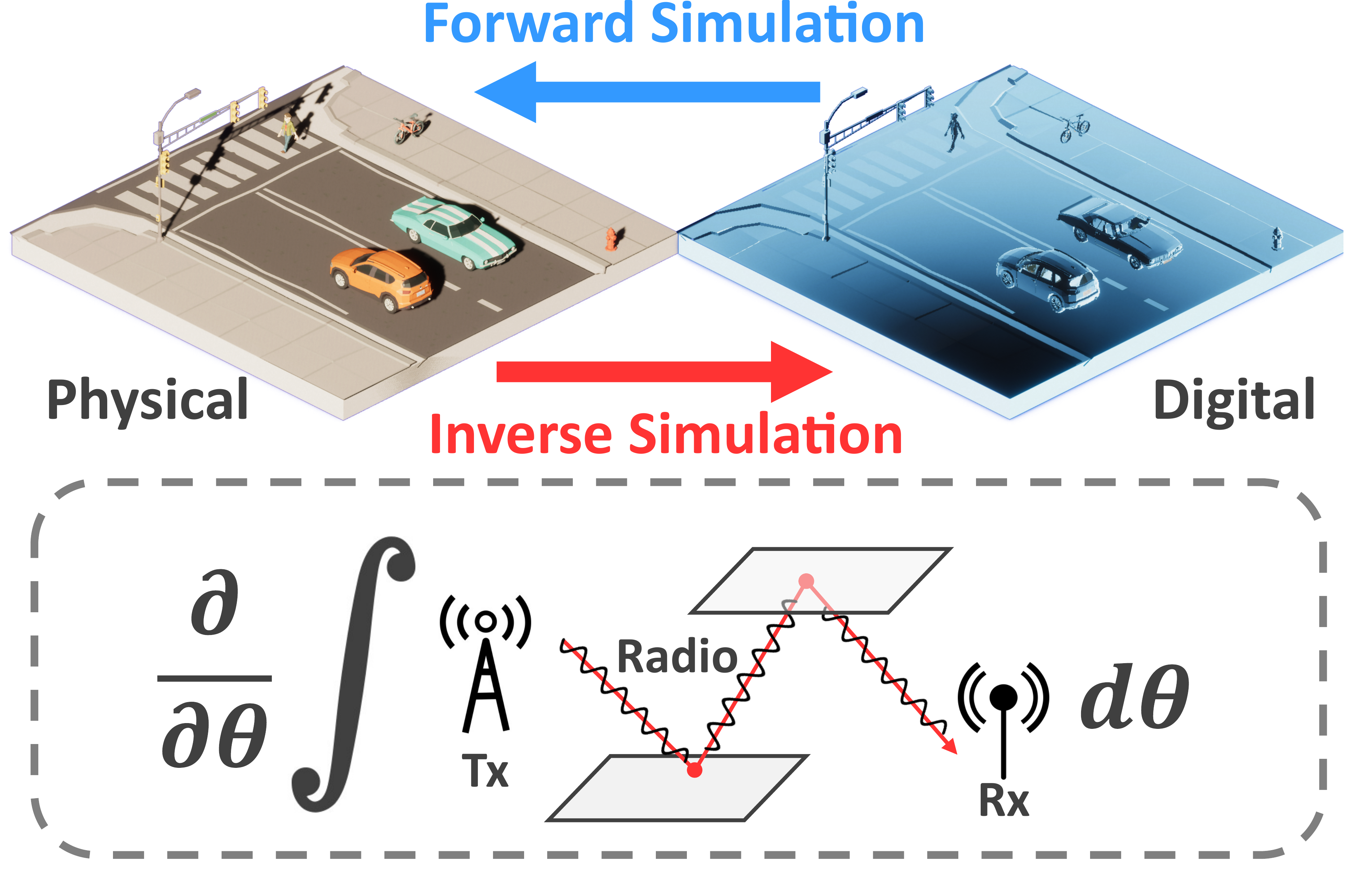}
\vspace{-19pt}
\caption{\name constructs digital twins by solving the inverse problem of RF simulation.}
\vspace{-12pt}
\label{fig:pull}
\end{figure}

To propel RF sensing to its next evolutionary phase, in this paper, we propose a novel framework that casts the problem of constructing digital twins (DTs) of objects in the physical world. 
As illustrated in Fig.~\ref{fig:pull}, a DT achieves high-fidelity simulation through continuous, bidirectional, and gradient-based interactions with its physical counterpart. This interaction distinguishes DT from unidirectional methods such as conventional RF ray tracing simulation (digital-to-physical) or sensing systems (physical-to-digital). 
Our DT framework, referred to as \textit{\textbf{\name}}, realizes such interaction by incorporating 
both in a closed-loop optimization framework. Unlike conventional forward simulation
techniques that use ray tracing to predict the RF signal patterns (output)
from physical twin's characteristic parameters (input), \name solves the \textit{inverse problem}, \ie, inferring the unknown input based on the observable output. 
Since the DT model is physics grounded and explainable, it overcomes the
reliance on training data or labels, and intrinsically handles \textit{unknown} 
scenes. On the other hand, its gradient-based optimization enables easy
integration of implicit knowledge and constraints (\eg, possible geometries
for given object categories) represented by pre-trained neural blocks.

To enable the closed-loop optimization in DT, the key challenge lies in 
\textit{differentiability} of the simulation steps that characterize how the RF signals propagate and interact with the physical objects. 
Differentiability allows for accurate computation of gradients between simulation outputs and the input parameters, facilitating efficient and robust optimization of the DT model. 
To meet this fundamental challenge, we identify and tackle two non-trivial subproblems.

\textit{(i) Discontinuity in complex simulation functions.} Traditional RF ray tracing simulations result in complex and often discontinuous functions, especially when the RF signals transition across or scatter at object boundaries. Inverting such simulations and making them differentiable presents non-trivial challenges. To overcome these challenges, we first identify and model the sources of discontinuity. We then introduce a path-space differentiation and reparameterization technique that enables differentiable simulation of the RF channel impulse response (CIR). This technique allows smooth gradient propagation through the simulation, thus enhancing the closed-loop optimization of the DT. 

\textit{(ii) Local non-convexity caused by the radio wave's periodic properties.} The periodic nature of RF signals introduces non-convexity in simulating the signal properties including strength and angles.  Direct application of gradient descent on RF signals, particularly the radar signals widely used in RF sensing, often fails to converge due to periodic local minima occurring at every wavelength.  To circumvent this issue, we incorporate a novel surrogate model for the radar range profile. This model not only generates smooth and accurate gradients, but also significantly reduces the computational overhead of the optimization framework. 

Harnessing the power of differentiability, our \name framework implements an iterative algorithm to optimize the DT's parameters, guided by the backpropagated gradients (Fig.~\ref{fig:pull}). 
\name allows for \textit{explicit} characterization of the DT using a comprehensive set of visual (3D geometry, position, pose, \etc) and RF (material property, radar configuration, \etc) parameters, which are explainable and can support diverse downstream tasks. In addition, \name can incorporate pre-trained neural networks as \textit{implicit knowledge} to constrain its optimization, enabling a hybrid model-driven and data-driven DT simulation.  

We have implemented the \name framework and conducted comprehensive experiments to verify its effectiveness. Our microbenchmark experiments demonstrate that \name accurately estimates gradients of the simulated signals \wrt various scene parameters. These precise gradients enable \name to solve complex inverse problems, such as 3D reconstruction and object shape estimation. Additionally, we conduct three novel case studies to showcase \name's versatility: test-time adaptation for machine learning systems, hybrid RF-visual sensing for robotics, and physically constrained parameter optimization. These case studies illustrate that \name can enable new capabilities for both data-driven and model-driven systems while enhancing their performance.

The key contributions of \name are three folds:

\noindent $\bullet$ 
We propose \name, a novel simulation-in-the-loop optimization framework that enables bidirectional, physics-based interaction between DTs and their physical counterparts. 
\noindent $\bullet$ 
We develop robust solutions to overcome the fundamental differentiability challenges in solving the inverse problem, expanding the boundaries of existing RF simulation and sensing domains.  
\noindent $\bullet$ 
We perform an extensive evaluation of \name through practical implementation and novel case studies. 

\section{Background and Related Work}

\subsection{RF Simulation Pipeline}
\label{sec:forward_sim}

RF propagation simulation has been extensively studied, with ray tracing serving as a fundamental technique.
Ray tracing begins with \textit{scene state}, represented by a set of \textit{scene parameters} $\theta$, characterizing the geometry and material properties of the RF scene. 
In practice, the scene is often discretized into tiny \textit{mesh} elements to facilitate computational analysis. 
The core process of ray tracing involves computing valid propagation paths, including reflections, refractions, and diffractions. Non-line-of-slight (NLOS) is modeled via penetrations on multilayered materials.  Fresnel coefficients are applied to calculate amplitude and phase changes at material interfaces.
The image method is often used for tracing the propagation paths, particularly for specular reflections. Material properties are then incorporated to determine electromagnetic interactions along valid paths.
For each path $i$, the RF simulation outputs its CIR, characterized by time delay $\tau_i$, amplitude $\alpha_i$, and phase shift $\phi_i$. The final CIR at the receiver is then expressed as a superposition of all the multipath CIR contributions. 
The simulated CIR can be further processed to obtain structured representations such as \textit{spatial spectrum}, \ie, 3D representations of signal strength in the space.
Note that, unlike full-wave electromagnetic simulation based on Maxwell's equations, RF ray tracing usually cannot accurately simulate the absolute phase of signals due to sophisticated object-wave interactions and RF hardware artifacts. However, it can represent the \textit{relative phase} between paths with reasonable accuracy, which indirectly affects the signal strength components in CIR through multipath combination.

\subsection{Related Work}
\textbf{RF sensing for object characterization and reconstruction.}
Substantial recent research has explored RF sensing for reconstructing 3D objects and characterizing their geometries and materials.  
The resolution of traditional signal processing methods is fundamentally limited by the radar antenna aperture and sampling bandwidth \cite{qian20203d}. 
Data-driven methods overcome the limitation by incorporating pre-trained knowledge of object characteristics  \cite{zhao2019through,guan2020through,xue2021mmmesh,ren20213d,jiang2020towards,Wang2022WiMesh,kong2022m3track,Xue2022M4esh}. For example, they can reconstruct human postures \cite{xue2021mmmesh} and vehicles \cite{guan2019high} by fitting sparse radar point cloud to learned meshes.  These approaches depend heavily on training data.  Emerging generative models can combine synthetic visual 3D meshes with RF simulation \cite{chen2023rfgenesis} to augment the radar training data, thus improving generalization. However, they still fall short at ``customization'', \ie, adapting to site-specific multipath environment or uncommon object properties.  In contrast, \name is a DT optimization framework that can fit the physical scene without training or fine-tuning.

\noindent 
\textbf{Inverse rendering and simulation of visual scenes.}
Inverse rendering techniques in computer graphics can retrieve the physical parameters of objects (\eg, 3D geometries) in a scene from a sparse set of pictures \cite{marschner1998inverse}. 
Early research relied on partial knowledge of the scene, \eg, lighting \cite{marschner1997inverse}, texture \cite{wei2008inverse}, and material \cite{chen2020invertible}. 
Neural rendering techniques can implicitly represent 3D scenes and synthesize novel views through volumetric ray marching \cite{mildenhall2020nerf}. These methods lack an explicit 3D representation and differentiability.
More recently developed differentiable renderers  \cite{kato2020differentiable, petersen2022gendr, loubet2019reparameterizing} approximate the traditional graphics pipeline by enabling gradient-based optimization of 3D representations like meshes and point clouds.  
Extending inverse rendering into the RF domain poses unique challenges.  The complexity of multipath reflections and diffractions, coupled with the inherent phase periodicity of RF signals, necessitates novel approaches. In \name, we conduct a rigorous analysis of these domain-specific problems and develop corresponding solutions to address the differentiability issues intrinsic to RF inverse rendering.

\textbf{Inverse and differentiable RF simulation.}
The differentiable computing of RF signals and their propagation field is in its nascent stage. Although a few works have attempted to employ optimization-based techniques in the RF domain, they do not touch upon the challenges of discontinuity and local non-convexity, and cannot be directly applied to RF sensing.

MetaWave \cite{chen2023metawave} represents an early attempt at using differentiable simulation to generate adversarial reflected for radar deception. However, its simulation simplifies the RF wave properties,  focusing solely on the signal strength of the first-order reflections. 
Furthermore, MetaWave's edge sampling method, which requires tracking all triangle edges in mesh representations, proves computationally 
prohibitive for complex scenes.  
The Sionna RF simulator \cite{sionna} supports the optimization of antenna gain patterns and reflection/scattering patterns, which are also subwavelength optimization problems. 
However, Sionna is mainly designed for modeling the RF communication channel distributions, and cannot deal with the aforementioned differentiability challenges unique to RF sensing systems. 
Sionna's gradient computation is fundamentally inadequate for RF sensing tasks.  
Although Sionna claims capabilities in estimating material properties to enhance channel estimation accuracy, it relies on simplified assumptions. Specifically, it treats the propagation path as fixed and estimates the reflection coefficient from the received signal, resulting in a simplified concave and continuous problem that does not fully capture the complexities of real RF environments.


\section{System Design}
\label{sec:design}

\subsection{Overview}
\label{sec:overview}
Our \name framework introduces a new paradigm for RF simulation and sensing by inverting the conventional forward process.  
Conventional RF simulation, as described in Sec.~\ref{sec:forward_sim}, follows a unidirectional path: scene states $\theta \rightarrow$ CIR $\rightarrow$ radar range profile $R(f) \rightarrow$ spatial spectrum $S_{spect}$, where the CIR is characterized by $(\tau,\alpha,\phi)$, \ie, the time delay, amplitude, and phase shift of all propagation paths between the transmitter and receiver. In contrast, \name aims to infer the scene states from observable RF data: radar spatial spectrum $S_{spect} \rightarrow$ scene states $\theta$. Due to the absence of a closed-form solution for direct inversion, we formulate the inversion as an optimization problem, achieved by calculating the partial derivatives to enable end-to-end backpropagation.

\begin{figure*}[t!]
\centering
\includegraphics[width=0.98\textwidth]{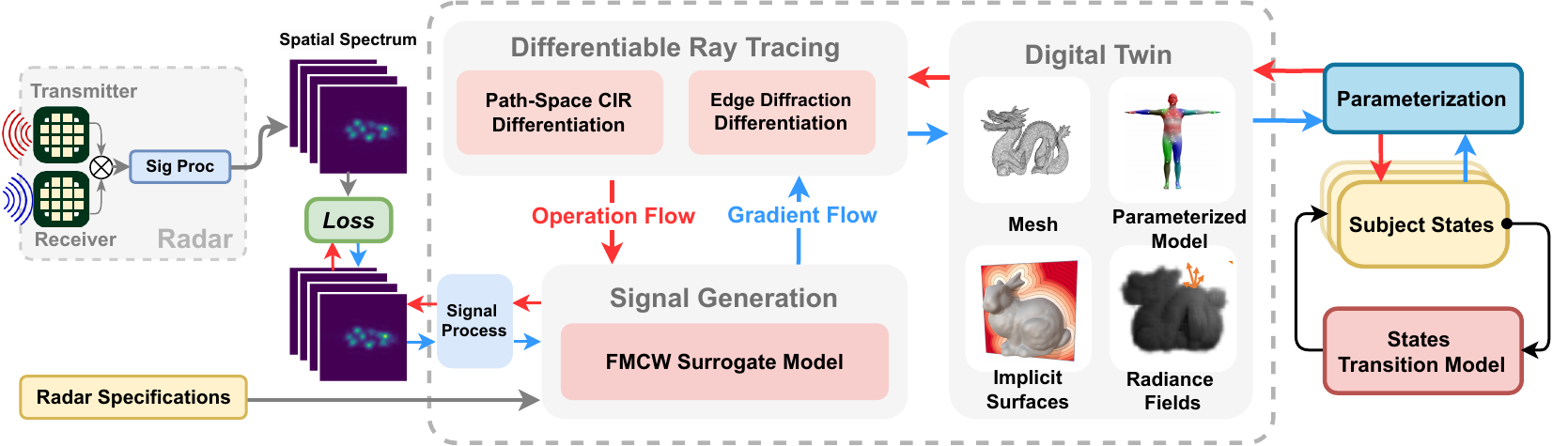}
\caption{ \name Workflow. }
\label{fig:framework}
\end{figure*}

\vspace{-5mm}
\begin{equation}
\small
\frac{\partial S_{spect}}{\partial \theta} = \sum_{i=0}^{N} ( \underbrace{\frac{\partial S_{spect}  }{\partial R(f)} }_{\text{Sec.~\ref{sec:diff_spect}} }\times \underbrace{\frac{\partial R(f)}{\partial (\tau,\alpha,\phi)}}_{\text{Sec.~\ref{sec:diff_fmcw}}} \times \underbrace{\frac{\partial (\tau_i,\alpha_i,\phi_i)}{\partial \theta}}_{\text{ Sec.~\ref{sec:diff_cir}}} ).
\end{equation} 

Our approach uses the chain rule to decompose the partial derivative into three subcomponents, each representing a key stage in the RF simulation process.    
The key challenge lies in ensuring the differentiability of these subcomponents while overcoming the local non-convexity.
To this end, \name incorporates two design components: Path Space Channel Impulse Response Differentiation and Radar Surrogate Model, as illustrated in Fig.~\ref{fig:framework}. The auxiliary components such as DT parameterization and optimization will be explained Sec.~\ref{sec:integration}.

\subsection{Challenges and Observations}
\label{sec:challenges}

\subsubsection{Discontinuity}
\label{sec:discontinuity}
RF propagation exhibits discontinuity primarily at two critical points: geometric edges and view-dependent edges.

\textbf{Geometric (Sharp) Edges}. A sharp edge is formed when two adjacent faces meet at an angle exceeding a specified threshold, creating a visible crease and wedge in the model. The sharp edge creates sudden changes in terms of the surface normal, and material causing discontinuity.  As described by the Geometric Theory of Diffraction (GTD), sharp edges introduce singularities in the field. 
These edges cause electromagnetic waves to scatter in multiple directions, creating a complex interference pattern. The diffracted field $E_d$ near a sharp edge can be approximated as $E_d \propto E_0 D(\theta,\phi)/\sqrt{kr}$, where $D(\theta,\phi)$ is the diffraction coefficient, creating a non-differentiable point at the sharp edge.

\textbf{View-dependent (Silhouette) Edges}. A silhouette edge is a line separating front-facing and back-facing polygons from the radar sensor's perspective, defining the outline of an object. 
These edges occur at the boundary between illuminated and occluded regions, where the line of sight between transmitter and receiver transitions from unobstructed to obstructed. This transition often results in a step function in the field strength, creating a non-differentiable point at the silhouette edge. In practice, the transition is rapid but not instantaneous due to diffraction effects, which can be modeled using the complementary error function: $E \propto E_0 \text{erfc}(\sqrt{2ik\rho})$, where $\rho$ is the distance from the silhouette edge. While this function is continuous, its steep gradient near the edge closely approximates non-differentiable behavior.

These non-differentiable points in RF propagation, whether from sharp edges or silhouette edges, introduce significant complexities in modeling RF environments, particularly in inverse problem formulations and optimization scenarios.

\subsubsection{Non-Convexity from Wave Periodicity and Finite Channel Samples}
\label{sec:NonConvexity}
The non-convexity in RF propagation primarily stems from the periodic nature of electromagnetic waves and the complex multipath environments. This non-convexity introduces significant challenges in convergence for optimization algorithms and inverse problems. The periodicity of the phase is a fundamental source of non-convexity. Path length ambiguity arises from the phase difference $\Delta\phi$ between two points, given by $\Delta\phi = 2\pi(\Delta L/\lambda)$, where $\Delta L$ is the path length difference and $\lambda$ is the wavelength. Due to the $2\pi$ periodicity, multiple path lengths can result in the same phase difference, causing ambiguity in optimization problems. Furthermore, standing wave points occur at locations where waves from different paths interfere destructively, creating a deep fading region with very low field strength. These points introduce sharp valleys in the optimization landscape, preventing gradient-based methods from convergence.

Employing wideband radar waveforms such as frequency-modulated continuous wave (FMCW), or orthogonal frequency division modulation (OFDM) \cite{sit2011ofdm} can often resolve the ambiguity 
Since electromagnetic waves with different wavelengths experience different phase shifts, it becomes less likely that different path lengths can result in the same phase difference for all the radar subchannels. 
Fourier analysis, \eg, discrete Fourier transforms (DFT), can be applied to the radar spectrum to separate the reflections with different path lengths.  
However, with a finite number of discrete channel samples, Fourier analysis inevitably leads to sidelobes~\cite{Oppenheim1998digital} whose peaks are equivalent to local minimums in the parameter space. This represents another critical source of non-convexity.

\begin{figure*}[t!]
\centering
\includegraphics[width=1\textwidth]{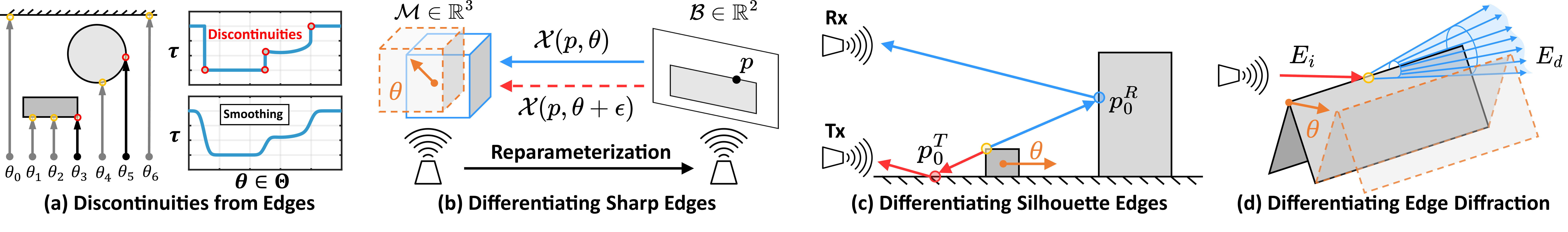}
\vspace{-15pt}
\caption{The rationale of mitigating discontinuities in RF simulation is to smooth the step and discontinuous function. However, smoothing while preserving physical accuracy in high-dimensional space is a non-trivial problem.  }
\vspace{-5pt}
\label{fig:path_space}
\end{figure*}
\subsection{Path-Space CIR Differentiation}
\label{sec:diff_cir}
\xya{In this section, we present our approach to addressing the discontinuity challenge, as categorized into two types in Sec.~\ref{sec:discontinuity}. We first formulate the discontinuous integral in path-space, and then handle geometric edges through reparameterization. We further design a custom ray tracing technique to track view-dependent edges. In addition, edge diffraction, often neglected in graphics but critical for RF applications, is treated as a combination of two types of discontinuity. 
}

The time-domain CIR, aggregating all the multipath components between the transmitter and receiver, can be represented as:
\begin{equation}
    \small
    h(t) = \int_{-\infty}^{\infty} a(\tau) \delta(t - \tau) e^{j\phi(\tau)} \text{d}\tau,
\label{eq:basic_ht}
\end{equation}
\noindent where $\delta$ is the Dirac delta function.
The corresponding frequency domain CIR can be obtained through the Fourier transform:

\begin{equation}
\small
H(f) = \int_{-\infty}^{\infty} h(t) e^{-j2\theta ft} \text{d}t.
\end{equation}

For further derivation, \textit{we transform Equ.~\eqref{eq:basic_ht} into path space by introducing a measurement contribution function} $f$ representing the amount of RF radiant power carried by the ray tracing paths $x = x_{\text{diffract}}\cup x_{\text{reflect}}$:

\begin{equation}
\small
    h(t) = \int_{\Omega} f(x) \, \text{d}\mu(x), \quad f(x) = x_\alpha \delta(t - x_\tau) e^{jx_\phi},
\label{eq:step1}
\end{equation}

\noindent where \(\Omega := \bigcup_{N=1}^{\infty} \mathcal{M}^{N+1}\) is the path space of all possible propagation paths, with each path \(x\) composed of key points \( (x_0, x_1, \dots, x_N) \) representing path intersection locations like reflection or refraction points. The measure \(d\mu(x)\) is defined as \(d\mu(x) := \prod_{n=0}^{N} dA(x_n)\), 
where \(dA(x_n)\) is the area element at each point \(x_n\), accounting for the geometric properties along the path. \(\mathcal{M}\) denotes the union of all object surfaces in the digital twin (DT) scene. 

Discontinuous integrands arise when a ray experiences occlusion or interacts with geometric edges. Consider an example DT scene with three geometries, as illustrated in Fig.~\ref{fig:path_space}(a). Suppose a parameter $\theta$ controls the translation along the X-axis, the relationship between $\theta$ and the time-delay $\tau$ is shown as a function in Fig.~\ref{fig:path_space}(a). 
The function is discontinuous at the edges of the geometries, leading to incorrect estimation of gradients.

\subsubsection{Path-Space Differentiation}
To correctly differentiate Equ.~\eqref{eq:step1}, we need to explicitly handle the boundary terms by tracking the speed of change of the domain boundary \wrt the target parameter following the Reynolds Transport Theorem \cite{reynolds1895iv}:

\begin{equation}
\small
\frac{\partial h(t)}{\partial \theta} = \underbrace{\int_{\Omega} \frac{\partial}{\partial \theta} f(x) \, \text{d}\mu(x)}_{\text{Interior}} + \underbrace{\int_{\partial \Omega} g(x) \, \text{d}\dot{\mu}(x)}_{\text{Boundary}},
\end{equation}

\noindent where $\partial \Omega$ is the union of all discontinuity points in the domain $\Omega$. Furthermore, \(\partial \Omega\) can be decomposed into two categories, sharp edges, and silhouette edges, \ie, \(\partial \Omega = \partial \Omega_{sharp} \cup \partial \Omega_{sil}\). 
The interior integral term is continuous \wrt the scene state $\theta$ and can be solved using a conventional automatic differentiation system. The boundary term, however, requires tracking all edges in the scene. A common solution is edge sampling, adopted in differentiable visual rendering and more recently in RF ray tracing \cite{chen2023metawave}. This method is costly, and its complexity increases drastically with scene complexity. Alternatively, biased heuristic approximation methods like those used in Sionna \cite{sionna} introduce biases that cause gradient error accumulation. This error accumulation becomes more pronounced as scene complexity increases, negatively affecting convergence on complex tasks.

\subsubsection{Differentiating Geometric Edges through Reparameterization}
\label{sec:reparam}
To overcome the limitations of edge sampling and biased approximation methods \cite{loubet2019reparameterizing}, we first parameterize all the geometries in the scene. 
Following previous studies in optical differentiation, continuum mechanics, and topology, all regular geometrical surfaces $\mathcal{M}$ in the scene can be considered to be deformed from some abstract 2D manifold $\mathcal{B}$, referred to as the \textit{reference surface}, as shown in Fig. \ref{fig:path_space}(b).
The deformations $\mathcal{X}(\cdot, \theta)$ are smooth functions that map reference surfaces $\mathcal{B}$ to regular surfaces $\mathcal{M}$, where $\theta \in \mathbb{R}$ is a controlling parameter. A regular surface controlled by parameter $\theta$ is referred to as an \textit{evolving surface} $\mathcal{M}(\theta)$.

We denote the propagation paths in reference space as \(\Omega_0\) and the boundary of the propagation path as \(\partial \Omega_0\). The key idea of our reparameterization technique is to construct the deformation
$\mathcal{X}$ such that \(\partial \Omega_{sharp} \in \Omega_0\), so that the discontinuity originates from sharp edges can be solved. \textit{This essentially works as a change of variables}. Such mapping $\mathcal{X}$ maps the propagation path $x = (x_0, \dots, x_N)$ to the reference space $p = (p_0, \dots, p_N)$ conditioned on mapping parameter \(\theta\). Applying such reparameterization yields a reformed CIR equation:

\vspace{-7pt}
\begin{align}
\small
h(t) = \int_{\Omega_{0}} \hat{f}(p) \, \text{d}\mu(p), \quad x = \mathcal{X}(p, \theta), \textrm{where}\\
    \hat{f}(p) = f(x) \prod_{n=0}^{N} J(p_n, \theta), \quad J(\mathbf{p}, \theta) := \left\| \frac{\text{d}A(\mathbf{x})}{\text{d}A(\mathbf{p})} \right\|,
\end{align}

\noindent and $J(\mathbf{p}, \theta)$ is the Jacobian matrix from the reparameterization. To this end, the final form of the path-space integral becomes:

\vspace{-7pt}
\begin{equation}
\small
\frac{\partial h(t)}{\partial \theta} = 
\underbrace{
\int_{\Omega_0} \frac{\partial}{\partial \theta} \hat{f}(p) \, \text{d}\mu(p)
}_{\text{Interior}}
+ 
\underbrace{
\int_{\partial \Omega_0} \hat{f}(p) V_{\partial}(p_K) \, \text{d}\dot{\mu}(p),
}_{\text{Boundary}}
\label{eq:reparameter_integral}
\end{equation}

\noindent where $\Omega_0$ is the propagation paths in reference space and $\partial\Omega_0$ is the boundary of the propagation path space.

\textit{The reparameterization essentially transforms sharp edges in regular surface \(\mathcal{M}\) to interior points in the reference surface \(\mathcal{B}\)}. Therefore, it resolves the discontinuity due to sharp edges. The remaining boundary term in Equ.~\eqref{eq:reparameter_integral} originates from the discontinuity due to view-dependent edges. Within the boundary term, $\hat{f}(p)$ represents the path contribution function. It integrates over the space of boundary paths $\partial\Omega$, where each path $p_K = (p_0, \ldots, p_N)$ contains a vertex $p_K$ situated precisely on a ``visibility'' boundary, defined as $p_K \in \Delta\mathcal{B}(p_{K-1})$. The function $V_\partial(p_K) = n_\partial(p_K) \cdot \frac{dp_K}{d\theta}$ quantifies the rate at which this boundary shifts \wrt scene state $\theta$, effectively measuring the sensitivity of RF signal occlusion. Here, $n_\partial(p_K)$ denotes the unit normal of the visibility boundary. 
Furthermore, we define the boundary segment caused by silhouette edges as the two-point-path pointing to a visibility boundary:
$(p_{K-1}, p_K) \in \mathcal{B}\times\Delta\mathcal{B}(p_{K-1}).$
    

\subsubsection{Differentiating View-dependent Edges}
\label{sec:view_dependent}
By reparameterization, we can accurately compute derivatives even in the presence of discontinuities. However, computing the boundary term in Equ.~\eqref{eq:reparameter_integral} requires searching for silhouette edges, 
which leads to a significant computational cost. Inspired by recent work \cite{Zhang:2020:PSDR}, we apply boundary path segmentation to avoid silhouette detection at every vertex. \xya{Essentially, the ray paths are created from the edges of the object controlled by $\theta$ expanding along two opposite directions to construct the path between Tx and Rx, as shown in Fig.~\ref{fig:path_space}(c).} We separate the boundary path into a transmission path \(\bar{p}^T\) and a receiver path \(\bar{p}^R\) at the discontinuity point \(p_K\): $\bar{p} = (p_t^T, \dots, p_0^T, p_0^R, \dots, p_r^R),$
where \((p_0^S, p_0^D) \in \mathcal{B}\times\Delta\mathcal{B}(p_{K-1})\). Then we can rewrite the boundary term of Equ.~\eqref{eq:reparameter_integral} as a combination of contributions from the source segment \(\hat{f}^S\), the boundary segment \(\hat{f}^B\) and the detection segment \(\hat{f}^D\):
\begin{equation}
\small
    \int_{\mathcal{B}}\int_{\Delta\mathcal{B}(p_{K-1})}\left(
    \int_{\Omega_0} \hat{f}^T\text{d}\mu^T
    \right)
    \hat{f}^B
    \left(
    \int_{\Omega_0} \hat{f}^R\text{d}\mu^R
    \right)
    \text{d}l(p_0^R)\text{d}A(p_0^T),
\label{eq:multi-direction-integral}
\end{equation}
where \(\hat{f}^T, \hat{f}^B, \hat{f}^R\) are all components of \(\hat{f}(p) V_{\partial}(p_K)\).

Through the multi-directional integral in Equ.~\eqref{eq:multi-direction-integral}, we can construct a boundary path as follows. Firstly, we sample the boundary segment \((p_0^T, p_0^R)\). Then we build the path from the source and the path towards the detector using 
bidirectional or unidirectional path sampling. 
This approach does not involve costly silhouette detection and therefore lowers the computational cost for the boundary term.

By combining this approach with reparameterization, we can bridge the gap between continuous and discontinuous changes in path space. 

\subsubsection{Differentiating Edge Diffraction}
\label{sec:dtd}

\xya{Simulating edge diffraction involves tracking wedges in the scene following the methods of the Uniform Theory of Diffraction (UTD). The method of differentiating edge diffraction can be treated as a combination of the methods proposed in Sec~\ref{sec:reparam} and \ref{sec:view_dependent}, as wedges are a subset of geometric edge and require tracking to construct diffraction path between Rx and Tx, as shown in Fig.~\ref{fig:path_space}(d).}
The diffracted field $E_d$ at a point can be expressed as $E_d = E_i D A(s) e^{-jks},$
where $E_i$ is the incident field, $D$ is the diffraction coefficient, $A(s)$ is the spreading factor, $k$ is the wavenumber, and $s$ is the distance from the diffraction point to the observation point.
\(\mathcal{E} \subseteq\partial\Omega\) denotes the edge points that have an edge diffraction effect. \(x_R\) is the position of the receiver. 
We can formulate the contribution of edge diffraction to 
the overall signal as:
\begin{equation}
\small
    h_D(t) = \int_{\mathcal{E}}E_d(\overline{xx_R})\text{d}A(x). 
\end{equation}
The differentiation of the diffraction process follows a similar approach including parameterization. Firstly, we map the edges to the reference surface \(\mathcal{B}\). Then we can calculate its derivative the same way as Equ.~\eqref{eq:reparameter_integral}. Since the edges \(\mathcal{E}\) are simple curves, the mapped points \(\mathcal{E}_0\) cannot be a plane and thus have no interior. Therefore, the derivative of the diffraction can be formulated as:
\begin{equation}
\small
    \frac{\partial h_D(t)}{\partial \theta} = \int_{\partial \mathcal{E}_0} \hat{E}_d(\overline{pp_R})V_{\partial}\text{d}l(p).
\end{equation}

\subsubsection{Monte Carlo Estimation of CIR Integrals}

In practical implementations, RF ray tracing simulators commonly estimate the CIR by integrating discrete propagation paths using the Monte Carlo method. \name simulates the CIR following the same approach, but must explicitly derive the gradients in this process. To this end, we note that the continuous integral of Equ.~\eqref{eq:step1} can be approximated by a sum over $N$ sampled paths:
\begin{equation}
\small
h(t) \approx \frac{1}{N} \sum_{i=1}^N \hat{f}(p_i),
\end{equation}
where $p_i$ are paths sampled from the path space $\Omega_0$. 

To estimate the derivative \wrt parameter $\theta$, we first apply the divergence theorem to convert the integral domain of the boundary term from \(\partial \Omega_0\) to \(\Omega_0\). The process can be formulated as:
\begin{equation}
\tiny
\begin{aligned}
    \int_{\partial\Omega_0}\hat{f}(p)V_{\partial}(p_K)\text{d}\dot{\mu}(p)
    =& \sum_{N=1}^\infty\sum_{k=0}^{N-1}\int_{\mathcal{B}^{N+1}}\left[\nabla\cdot\left(\hat{f}_kV_{\partial}\right)\right](p_k)\text{d}\mu(p) \nonumber\\
    =& \int_{\Omega_0}\sum_{k=0}^{N-1}\left[\nabla\cdot\left(\hat{f}_kV_{\partial}\right)\right](p_k)\text{d}\mu(p),
\end{aligned}
\end{equation}
Then we can calculate the derivative using Monte Carlo estimation:
\begin{equation}
\small
\frac{\partial h(t)}{\partial \theta} \approx \frac{1}{N_I} \sum_{i=1}^{N_I} \left(\frac{\partial}{\partial \theta}\hat{f}(p_i) + \sum_{k=0}^{N-1} \left[\nabla\cdot\left(\hat{f}_kV_{\partial}\right)\right](p_k)\right).
\end{equation}

This approach allows for efficient estimation of both the CIR and its derivatives. It is worth noting that in practical implementation, the multipath components of CIR are calculated separately, and hence the partial derivative $\frac{\partial \hat{\tau}}{\partial \theta}$, $\frac{\partial \hat{\alpha}}{\partial \theta}$,  $\frac{\partial \hat{\phi}}{\partial \theta}$ can be estimated via:

\vspace{-4mm}
\begin{equation}
\small
\left(\frac{\partial \tau_i}{\partial \theta},
\frac{\partial \phi_i}{\partial \theta},
\frac{\partial \alpha_i}{\partial \theta}\right)=
\frac{\partial}{\partial \theta} \hat{f}_y(p_i) + \hat{f}_y(p_i)V_\partial(p_{K,i}), \ y \in \{\tau, \phi, \alpha\}.
\end{equation}
\vspace{-4mm}

With this measure, the time-delay ($\tau$) and amplitude ($\alpha$) components of the CIR become differentiable. Consequently, we can backpropagate the gradients from the CIR to the 
DT scene ($\mathbf{\theta}$). \xya{Whereas the calculation of the partial derivatives is now complete, it may still suffer from local non-convexity issues (Sec.~\ref{sec:challenges}). We now introduce the surrogate models tailored to mitigate these challenges.}

\subsection{Radar Surrogate Model}
\label{sec:surrogate}

\xya{Radar systems utilize modulated signals to determine the range and angle of reflections, both of which face the non-convexity challenges (Sec.~\ref{sec:NonConvexity}). 
As illustrated in Fig.~\ref{fig:periodic}, the conventional radar signal processing methods exhibit oscillations and local non-convexities that hinder gradient-based optimization techniques. To address the issue of local non-convexity, we propose \textit{surrogate models that decouple phase information from range and angle calculations, resulting in a smoother loss landscape}. It is important to note that this surrogate model is employed solely in the simulation; the actual signal processing still relies on conventional methods. In other words, rather than simulating and outputting a raw IF signal that is subsequently processed by the same algorithms used in real radar systems, our simulation can choose to directly output a range profile or spatial spectrum to benefit convergence. This task is non-trivial as it requires the simulation to closely mimic real signals, including the sidelobes and wave interference effects (Sec.~\ref{sec:NonConvexity}), to preserve physical accuracy.}

\subsubsection{Radar Waveform Surrogate}
\label{sec:diff_fmcw}

\begin{figure}[htb!]
\centering
\vspace{-10pt}
\includegraphics[width=0.48\textwidth]{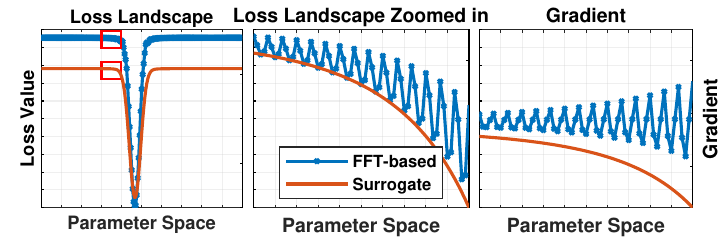}
\vspace{-5pt}
\caption{ Periodic local minima in the loss landscape occur when optimizing on radar signals. The optimization aims to find the global minimum $\theta \in \mathbb{R}$ in terms of the loss (y-axis) via gradients. }
\label{fig:periodic}
\end{figure}

Our surrogate model primarily tackles FMCW radar signals, though the same challenges and model are applicable to other modulation schemes. The radar received signal is based on the CIR components $(\tau, \alpha, \phi)$:

\vspace{-4mm}
\begin{equation} \label{eq:fmcw_if}
\small
    S_{\text{IF}}(t) = \sum_{i=1}^{N} a_i \exp\left(2\pi j \left(\frac{B}{T} \tau_i t + f_0 \tau_i - \frac{B}{2T} \tau_i^2 + \phi_i\right)\right),
\end{equation}
\vspace{-3mm}

The complex-valued radar range profile can be calculated via FFT:

\vspace{-3mm}
\begin{equation} \label{eq:fmcw_fft}
\small
    R(f) = \sum_{n=0}^{N-1}(S_{\text{IF}}[n] \cdot w[n]) \cdot \exp\left(-j \frac{2\pi}{N} f n\right), 
\end{equation}
\vspace{-3mm}

\noindent where $w[\cdot]$ is the hamming window function used to reduce the side lobes in the frequency domain when performing FFT. 
The range profile is the foundation for radar signal processing. It can be used to derive the point cloud \cite{prabhakara2023radarhd}, the spatial spectrum \cite{guan2020through} or used as input features for data-driven models \cite{xue2021mmmesh}.  
However, such calculations suffer from severe convergence issues as described in Sec.~\ref{sec:challenges}  and 
Fig.~\ref{fig:periodic}, as well as heavy computation due to complicated calculation of both the time domain and frequency and the periodic phase function.

To circumvent this convergence issue, we propose a novel algorithm as a surrogate model for radar simulation, which directly constructs CIRs from valid rays and still retains the differentiability between them.  
Specifically, we approximate the radar range profile using a surrogate model $R(f)^s$:

\vspace{-3mm}
\begin{equation}
    \tiny
R(f)^s = \sum_{i} \underbrace{\frac{\exp\left(-\frac{(\tau_i - D(f))^2}{2\sigma^2}\right)}{\sum_{j} \exp\left(-\frac{(\tau_i - D(f_j))^2}{2\sigma^2}\right)}}_{G_i} \cdot \alpha_i  \cdot e^{j\phi_i},
\end{equation}
\vspace{-3mm}

\noindent where $P_i$ is a modified Gaussian kernel. $D(f) = \frac{c \cdot f}{2\cdot B / T}$ calculates the range bin from the given frequency.  
The surrogate model approximates the contribution of each path with a Gaussian pulse, with $\sigma$ controlling the spanning of the pulse width.

The $R(f)^s$ can be effectively differentiated. Here we provide the symbolic solution to the partial differentiation function for each CIR component:

\vspace{-3mm}
\begin{equation}
    \tiny
    \frac{\partial G_i}{\partial \tau_i} = \frac{\partial}{\partial \tau_i} \left( \frac{\exp\left(-\frac{(\tau_i - D(f))^2}{2\sigma^2}\right)}{\sum_{j} \exp\left(-\frac{(\tau_j - D(f_j))^2}{2\sigma^2}\right)} \right) \\
    = G_i \cdot \frac{(D(f) - \tau_i)}{\sigma^2} \cdot (1 - G_i),
\end{equation}
\vspace{-3mm}

\noindent where the partial derivative of the Gaussian kernels $\frac{\partial G_i}{\partial \tau_i}$ is first derived.  Then we can derive the derivative of \(\tau_i, \alpha_i, \phi_i\) on the surrogate model \(R(f)^s\).

\vspace{-3mm}
\begin{equation}
\small
    \frac{\partial R(f)^s}{\partial \tau_i} = \frac{\partial G_i}{\partial \tau_i} \cdot \alpha_i \cdot e^{j\phi_i}, \quad
    \frac{\partial R(f)^s}{\partial \alpha_i} = G_i \cdot e^{j\phi_i}, \quad
    \frac{\partial R(f)^s}{\partial \phi_i} = j \cdot G_i \cdot \alpha_i \cdot e^{j\phi_i}.
\end{equation}
\vspace{-3mm}

\noindent To this end, the partial derives for complex FMCW signal \wrt the CIR components is derived.

\subsubsection{Spatial Spectrum Surrogate}
\label{sec:diff_spect}

Mainstream spatial spectrum techniques for MIMO radar, such as beamforming and MUSIC, are differentiable \wrt the range profile (detailed in the anonymous supplementary material due to space constraints). This allows for the direct calculation of the partial derivative component, $\frac{\partial S_{spect}}{\partial R(f)}$, via automatic differentiation. However, these algorithms can be prone to local non-convexity issues due to 
phase interference. To address this, 
we modify the range surrogate model proposed in 
Sec.~\ref{sec:diff_fmcw} to a spatial surrogate by incorporating Airy disc functions which are widely used to describe 2D sidelobe patterns:
\begin{equation} 
\small 
E(\psi) = E_0 \left( \frac{2J_1(k r \sin \psi)}{k r \sin \psi} \right)^2 e^{j\phi}, 
\end{equation}

\noindent where $\psi$ represents the angle of observation, $k = 2\pi / \lambda$ is the wavenumber. $r$ denotes the radius of the aperture, and $J_1$ is the Bessel function of the first kind of order one. 
The phase $\phi$ is randomly assigned a value between $0$ and $2\pi$ to mimic the stochastic nature of phase variations. 
This adjustment allows for interference among paths with different phases while 
decoupling them from the time delay, effectively 
functioning as a random mask or dropout layer to aid convergence.

\section{System Integration}
\label{sec:integration}

\begin{figure}[t]
\centering
\includegraphics[width=0.5\textwidth]{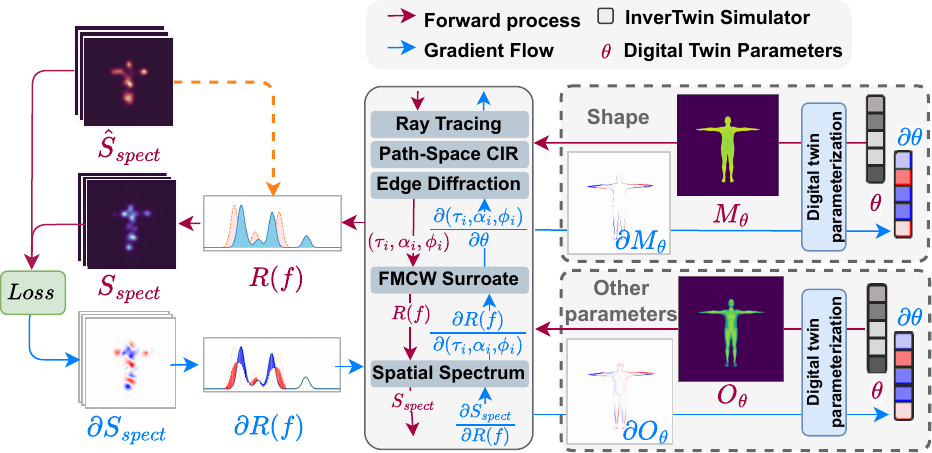}
\vspace{-10pt}
\caption{ \name forward simulation and backward propagation flow. }
\label{fig:gradient_flow}
\end{figure}

With the path-space differentiation and radar surrogate model, we can derive gradients from each step of the RF simulation, and propagate them back to the input (DT parameters) to form an optimization loop. We now elaborate on this integrated design following the end-to-end optimization workflow in Fig.~\ref{fig:gradient_flow}.

\subsection{Digital Twin Parameterization}
The DT is represented by a set of parameters $\Theta$, which can be categorized into different subsets depending on downstream applications.
Examples include, but are not limited to: \textit{(i)} 3D shape/geometry $\mathcal{M}_{\theta}$; \textit{(ii)} linear transformations, such as translation, rotation, and scaling; \textit{(iii)} scalar values like reflection patterns and dielectric constants which can be mapped to material distributions. 

Parameters $\theta$ that govern the geometry $\mathcal{M}_{\theta}$ are most critical for 3D reconstruction.
\name supports a wide range of differentiable geometry representations, including meshes, Signed Distance Fields (SDFs) \cite{park2019deepsdf}, SMPL model \cite{loper2023smpl}, and radiance fields \cite{mildenhall2021nerf}. While these representations remain an active area of research, \name is compatible with any representation that allows differentiation of ray intersections \wrt its control parameters.

\subsection{Digital Twin Optimization }
\label{sec:DTopt}

We formulate the interaction between the DTs and physical entities as an optimization process shown in Fig.~\ref{fig:gradient_flow}.
This process iteratively refines the initial DT parameters $\bm{\theta}_0$, leading to more and more accurate representation $\bm{\theta}^*$ of the real-world scene from the RF signals.  
The problem can be expressed mathematically as: 

\vspace{-4mm}
\begin{equation}
\small
\bm{\theta}^* = \arg\min_{\bm{\theta} \in \Theta} \mathcal{L}(S(\bm{\theta}), \bm{y}), \quad   \frac{\partial \mathcal{L}}{\partial \theta} = \frac{ \partial \mathcal{L} }{\partial P_{spect}} \times \frac{\partial S_{spect}}{\partial \theta},
\end{equation}

\noindent where $\mathcal{L}(\cdot, \cdot)$ is the loss function measuring the discrepancy between simulated and observed RF signals. We employ multiscale MSE as the loss function which 
captures the spectrum structure at different levels of resolution to mitigate the sparsity characteristics of the RF spectrum.  

To solve the optimization problem, we use Stochastic Gradient Descent (SGD) with regularization. Specifically, the iterative update is given by: $\bm{\theta}_{t+1} \leftarrow \bm{\theta}_{t} - \eta \left( \frac{\partial \mathcal{L}(S(\bm{\theta}), \bm{y})}{\partial \bm{\theta}} + \lambda \mathbf{L} \bm{\theta} \right),$
where $\eta > 0$ is the learning rate at iteration $t$. $\mathbf{L} \in \mathbb{R}^{n \times n}$ is a sparse, symmetric positive definite Laplacian matrix that discretizes Dirichlet energy \cite{solomon2014laplace}, and $\lambda$ is a weight controlling the strength of regularization.

Using regularized SGD, we iteratively refine the DT by sampling subsets of the RF signals, computing the discrepancy gradients updating the DT parameters. Upon convergence, 
the RF simulation output should closely match the observed/received RF signals. Meanwhile, \name yields an accurate parameterized replica $\bm{\theta}^*$ of the physical scene. 

A key advantage of \name lies in its ability to integrate pre-trained cross-domain knowledge to enhance the DT reconstruction.  Instead of searching the entire parameter space $\Theta$, even coarse-grained knowledge can impose additional constraints, limiting the search to a much smaller subspace.  As an example, for automotive perception applications, the reconstruction can be limited to ``roadway objects'', by utilizing pre-trained object shape decoders from datasets such as ShapeNet \cite{shapenet2015}. Since the exact object shape is unknown and can vary significantly, we use dynamic meshes to parameterize their 3D geometries.  We define a 3D density field based on a target resolution (\eg, 512 sampling points per dimension). To ensure computational efficiency, we use ShapeNet to compress the density field to a smaller parameter set, decompress them back to the desired $512\times512\times512$ using ShapeNet, and finally triangulate the density field to a mesh representation of the DT.  Unlike other end-to-end data-driven RF sensing methods, this approach decouples the pre-trained computer graphics knowledge from RF modeling, eliminating the need for laborious RF training data collection.

\begin{figure*}[ht!]
    \begin{minipage}{0.68\linewidth}
        \centering
        \setlength{\abovecaptionskip}{0pt}
        \includegraphics[width=1\linewidth]{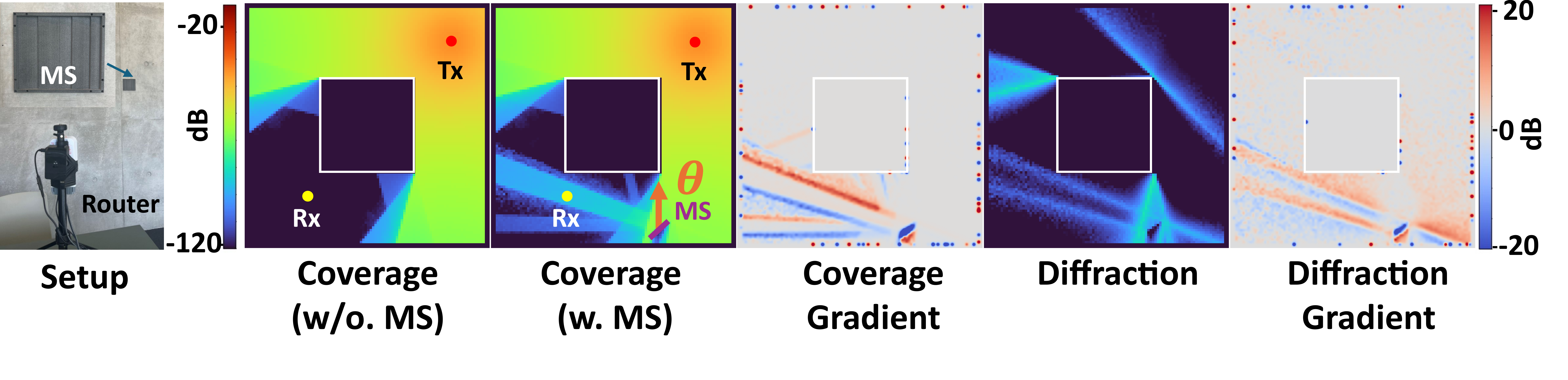}
        \caption{\name estimates the gradient of the coverage map \wrt MS translations $\theta$ for optimization, a capability that no existing simulator offers.}
        \label{fig:ms}
    \end{minipage}
    \hspace{2pt}
    \begin{minipage}{0.3\linewidth}
        \setlength{\abovecaptionskip}{0pt}
        \includegraphics[width=1\linewidth]{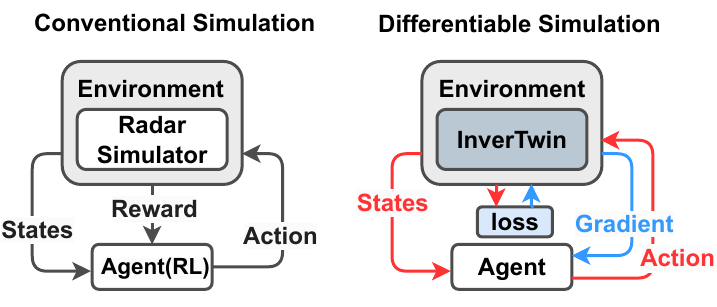}
        \caption{Differentiable Robotic simulation enables end-to-end training, step beyond conventional reinforcement learning design.}
        \label{fig:robot}
    \end{minipage}
\end{figure*}

\subsection{Software Implementation}
We implement the \name forward simulation process (Fig.~\ref{fig:gradient_flow}) using C++ and CUDA. Furthermore, we implement the \name optimization flow in PyTorch, which seamlessly incorporates the forward simulator. The ray tracing simulation is powered by hardware acceleration through NVIDIA's OptiX library \cite{parker2010optix} utilizing RT cores \cite{nvidia_ray_tracing}. The differentiation of the ray tracing model is primarily handled via automatic differentiation using the Dr.JIT library \cite{jakob2022dr}, which generates differential functions for each module during the forward simulation process at runtime. However, the auto-generated functions may not be optimal \cite{bangaru2023slang}, resulting in a large computation graph that becomes a bottleneck in the back propagation. Therefore, we manually implement the backward functions for path-space CIR differentiation (Sec.~\ref{sec:diff_cir}) and the surrogate model (Sec.~\ref{sec:surrogate}) to optimize its efficiency.

\section{Micro Benchmark Evaluation}

\begin{figure}[htb]
\centering
\includegraphics[width=0.49\textwidth]{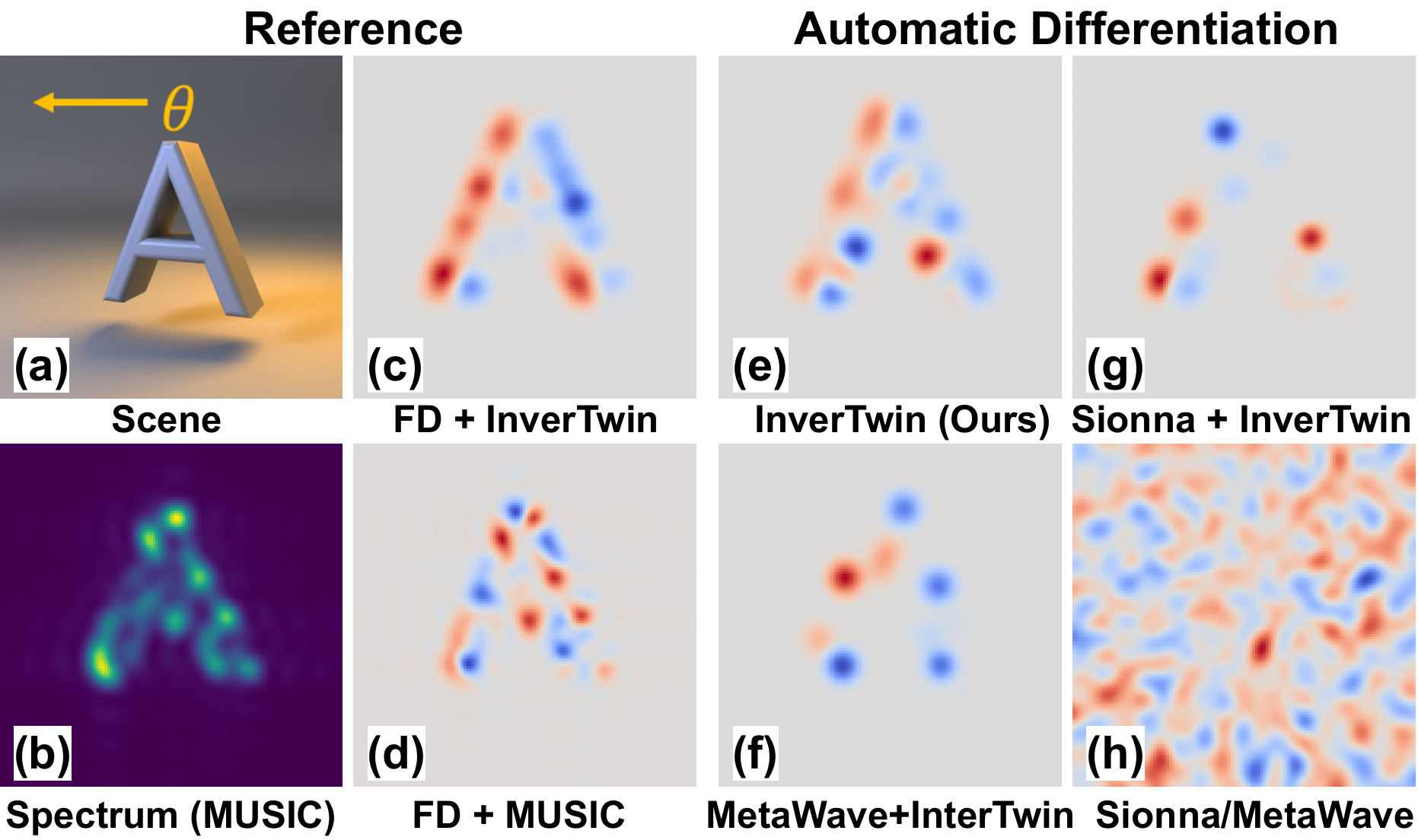}
\vspace{-5pt}
\caption{Gradient estimation results illustrating the effect of the parameter $\theta$ on the horizontal translation of a 3D letter.  Red regions indicate an increase in spectrum pixel values with $\theta$, while blue regions indicate a decrease.}
\label{fig:grad}
\end{figure}

\subsection{Gradient Estimation Evaluation}

We evaluate the effectiveness of \name in gradient estimation by benchmarking it against the widely used finite difference (FD) method, the standard for assessing differential renderers and simulations \cite{leveque1998finite, loubet2019reparameterizing, li2018differentiable}. The FD method approximates the gradient using $ \frac{\partial f}{\partial x} \approx \frac{f(x + \epsilon) - f(x)}{\epsilon}$, where a small perturbation $\epsilon$ is added to a single parameter, and the difference in the resulting simulations is calculated. While accurate, FD can only estimate one gradient parameter per simulation round, whereas real-world scenarios often require the simultaneous backpropagation of thousands of parameters. Consequently, FD is typically employed in small controlled simulation environments to obtain the ground truth gradient. 

For our evaluation, we use a 3D letter ``A'' model with a target parameter $\theta$ controlling its horizontal translation, as shown in the scene in Fig.~\ref{fig:grad}a.  We use Multiple Signal Classification (MUSIC) \cite{schmidt1986multiple} as a comparison, as it demonstrates promising results in generating MIMO radar spatial spectrum. \name estimates the gradient for each voxel within the spatial spectrum relative to the scene parameters. The projected 2D views of voxel gradients are shown in Fig.~\ref{fig:grad}e, in comparison with baselines.
The FD method applied to the MUSIC algorithm still 
exhibits a periodic pattern along edges , unsuitable for optimization. Direct differentiation of the baseline simulation on MUSIC yields corrupted gradients (Fig.~\ref{fig:grad}d). However, with our radar surrogate model design (Sec.~\ref{sec:surrogate}), FD's results provide a gradient that directly reflects object translation (Fig.~\ref{fig:grad}c). 
We use the FD method as the ground truth.
Even with the surrogate model to overcome nonconvexity, Metawave and Sionna (Fig.~\ref{fig:grad}g,h) fail to accurately capture gradients, particularly around edges, achieving an MAE of 0.63 and 0.69, respectively. Our path space differentiation approach (Sec.~\ref{sec:diff_cir}) explicitly handles the edge discontinuities, allowing \name to preserve critical details of the scene, decreasing the MAE to 0.59. 

\begin{figure}[ht!]
\centering
\includegraphics[width=0.48\textwidth]{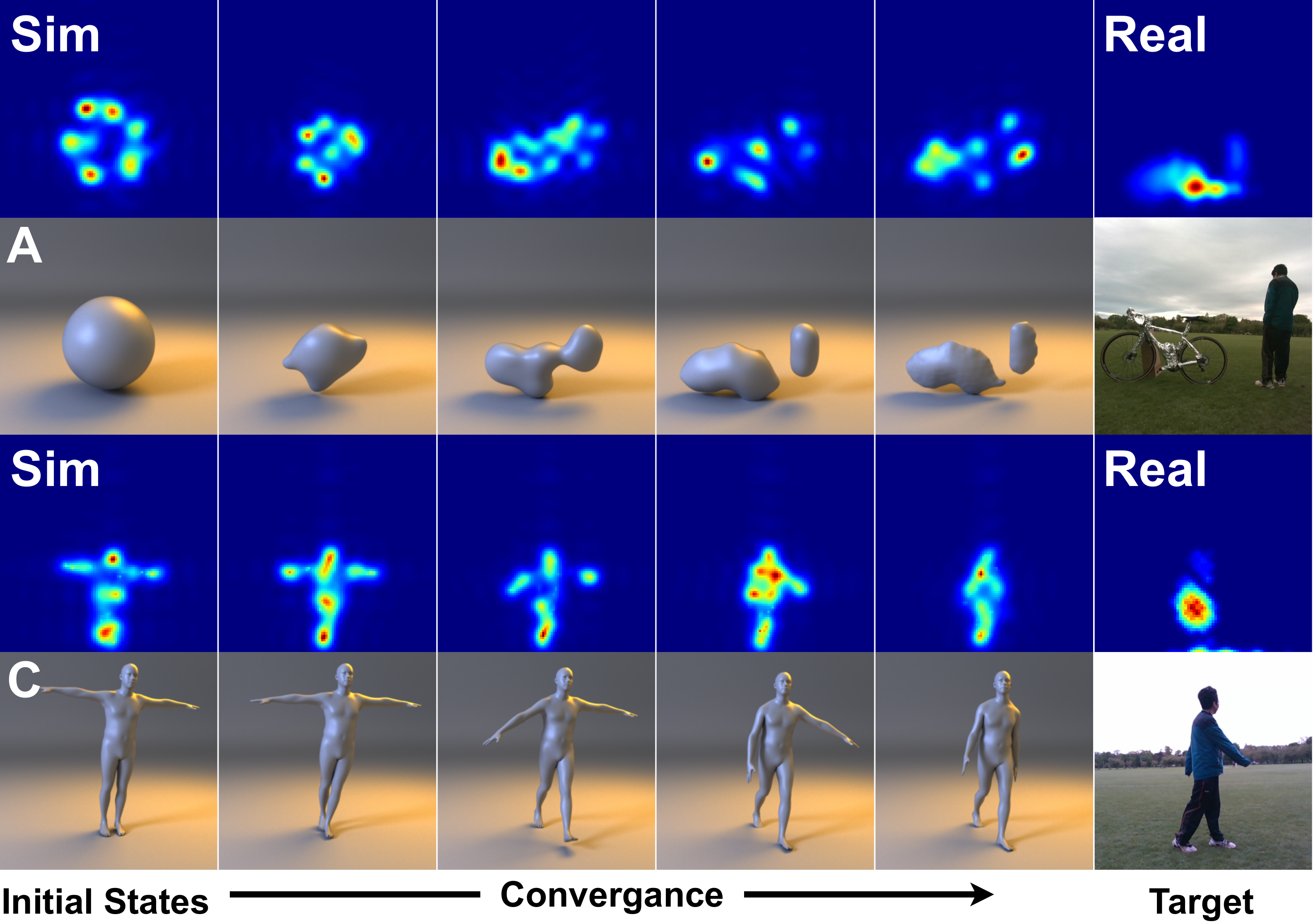}
\vspace{-10pt}
\caption{Convergence of the DT optimization without and with pre-trained knowledge of ``possible shapes of vehicles''.}
\label{fig:opt}
\vspace{-5pt}
\end{figure}

\vspace{-5pt}
\section{Discussion and Future Works}
\noindent\textbf{Comparison and integration with fitting-based methods}.
\name shares similarities with conventional fitting-based methods, such as point cloud registration, that iterative adjust the scene representation to minimize the difference between geometry and sparse observations (\eg, point clouds or spatial spectrum). The key difference lies in how \name incorporates physical laws governing how the observations are generated. These laws include how RF signals respond to material properties, specular reflection, diffraction, \etc. 
As such, \name can augment the traditional techniques. For example, it can use their outputs as an initial guess of the DT scene, and further refine the scene through its optimization framework.   

\noindent \textbf{End-to-end robotic training}.
Differentiable simulation for robotics is an emerging research area that aims to combine differentiable computing with physics-based simulation. It seeks to create simulation-in-the-loop training of robotic control policies.   
\name can be used as a differentiable simulation engine when the robot employs radar perception to facilitate decision making, as shown in Fig.~\ref{fig:robot}. 
Unlike traditional simulation methods, this approach allows gradient information to propagate directly from the simulation environment to the control policy. It can enable full end-to-end training of the robotic system to accelerate the learning process and improve performance.

\vspace{5pt}
\section{Conclusion}

We have introduced the design of \name, the first fully differentiable simulation system that solves the inverse problem of RF simulation in order to reconstruct digital twins using RF signals. By efficiently estimating the gradients of simulated signals \wrt scene parameters, \name uniquely deduces simulation inputs from outputs. To ensure reliable convergence, we developed several advanced techniques, including path-space CIR differentiation and radar waveform surrogate models. Through demonstrations across diverse use cases within one unified framework, we envision \name as a pioneering step, opening new research directions in computational RF for future studies.

\newpage
\bibliographystyle{plain}
\bibliography{sigproc}

\end{document}